\documentclass[aps,prb,reprint,showpacs,superscriptaddress]{revtex4-1}

\usepackage{amsbsy,amsmath,amssymb}
\usepackage{textcomp,color}
\usepackage{graphicx}
\usepackage{bm}
\usepackage{times,xspace}
\usepackage{ulem}

\begin{document}


\title{%
  Dimensional crossover in the quasi-two-dimensional Ising-O(3) model
}


\author{Y. Kamiya}
\affiliation{%
  Theoretical Division, Los Alamos National Laboratory, 
  Los Alamos, New Mexico 87545, USA
}%

\author{N. Kawashima}
\affiliation{%
  Institute for Solid State Physics, University of Tokyo, 
  Kashiwa, Chiba 227-8581, Japan
}%

\author{C. D. Batista}
\affiliation{%
  Theoretical Division, Los Alamos National Laboratory, 
  Los Alamos, New Mexico 87545, USA
}%


\date{\today}

\begin{abstract}
  We present the results of our Monte Carlo simulation of the Ising-O(3) model on two-dimensional (2D) and quasi-2D lattices.
  This model is an effective classical model for the stacked square-lattice $J_{1}$-$J_{2}$ Heisenberg model, 
  where the nearest-neighbor ($J_1$) and next-nearest-neighbor ($J_2$) couplings are frustrated and we assume that $J_2$ is dominant.
  We find an Ising ordered phase in which the O(3) spins remain disordered in a moderate quasi-2D region. 
  There is a single first-order transition for a sufficiently large 3D coupling, 
  in agreement with a renormalization group treatment.
  The subtle region in which the single transition splits into two transitions is also discussed and compared against recent measurements of two very close transitions in BaFe$_2$As$_2$.
  Our results can provide a qualitative explanation of the experiments on ferropnictides, namely the observed sequence and orders of the structural and magnetic transitions, in terms of the ratio between the inter-layer and intra-layer coupling.
\end{abstract}


\maketitle


\section{%
  \label{sec:introduction}
  Introduction
}
The dimensional crossover [from one-dimensional (1D) or 2D phenomena to 3D ones] has been an important issue in condensed matter physics because there are several quasi-low-dimensional materials that exhibit complex behaviors. 
Enhanced quantum fluctuations in low-dimensional systems can give rise to novel quantum states of matter.
Although a small 3D coupling always exists in real systems and ordered phases are usually stabilized at low temperatures, 
the dimensional crossover is still governed by the low-dimensional physics.

\begin{figure}[!t]
  \begin{center}
    \includegraphics[angle=90,width=19.5pc, bb=137 96 469 693, clip]{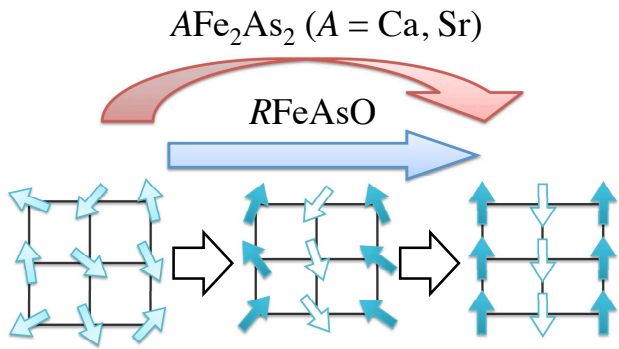}
    \vspace*{-0.10cm}
    \caption{%
      (Color online)
      Schematic pictures of the sequence of the transition(s) observed in the parent compounds of the iron-based superconductors.
      Shown are the  disordered state with tetragonal lattice symmetry at high $T$'s (left), 
      the Ising ordered state with broken lattice symmetry and \textit{short-range} magnetic ordering at intermediate $T$ (center), 
      and the lowest-$T$ state with broken lattice symmetry and long-range stripe-like magnetic ordering (right).
      The symmetry also allows for the other lattice distortion obtained by rotating the presented lattice by $\pi/2$.
      \label{fig:shematic-pic}
    }
  \end{center}
  \vspace*{-0.35cm}
\end{figure}

When a system combines fluctuating {\it continuous and discrete} degrees of freedom, 
the interplay between them leads to an even richer dimensional crossover. 
The simple reason is that they respond to a weak 3D coupling in a qualitatively different way. 
This physics has recently attracted  particular interest in the context of  parent compounds of  iron-based superconductors.~\cite{Paglione2010high,Lumsden2010magnetism,Ishida2009to} 
Among the iron-based compounds, the 1111-type quasi-2D materials $R${Fe}{As}{O} ($R$ is a rare earth ion)~\cite{Cruz2008EXP:magnetic,McGuire2008phase} and
the 122-type 3D compounds $A${Fe}$_2${As}$_2$ ($A$ is an alkaline earth ion)~\cite{Krellner2008magnetic,Rotter2008Spin-density-wave,Goldman2008Lattice,Yan2008Structural} 
constitute a subgroup with the following low-energy properties:~\cite{Paglione2010high}
(i) they are metallic, (ii) they undergo a tetragonal-orthorhombic structural transition, and (iii) they stabilize a stripe-like spin-density-wave (SDW) order at a wavevector $(\pi, 0)$ at the lowest temperatures.
The structural transition is an Ising-like transition in the sense that it breaks a $Z_2$ spatial symmetry.
The continuous and discrete characters of these two broken symmetries lead to qualitatively different magnetic and structural fluctuations whose interplay should be strongly affected by the magnitude of the 3D coupling.
Indeed, experiments have confirmed that the structural and magnetic transitions take place simultaneously via a first-order transition in most of the nearly 3D undoped 122 compounds,~\cite{Krellner2008magnetic,Goldman2008Lattice,Yan2008Structural} whereas the lattice distortion occurs at a slightly higher temperature in the more quasi-2D 1111 materials (Fig.~\ref{fig:shematic-pic}).~\cite{McGuire2008phase} 
In the latter case, both  transitions seem to be of second order or at least very weakly first order.~\cite{McGuire2008phase} 
The proximity between the two transitions suggests that the magnetic ordering plays a central role in the lattice distortion.~\cite{Fang2008theory,Xu2008Ising}
Previous density functional studies also indicate that the structural distortion may be driven by the interaction between magnetic degrees of freedom.~\cite{Yildrim2008origin}

These qualitative and ubiquitous features of the iron pnictides are expected to be universal in the sense that they should only depend on symmetry, dimensionality, 
number of components of the order parameter, and range of interactions. 
For this reason, it has been argued that the $J_{1}$-$J_{2}$ Heisenberg model~\cite{Misguich2004REVIEW} 
is very useful in spite of its simplicity as a purely local spin model.~\cite{Fang2008theory,Xu2008Ising}
Here, we consider this model on a stacked square lattice. 
The model includes  in-plane nearest-neighbor ($J_1$) and next-nearest-neighbor ($J_2$) exchange couplings that lead to a high degree of frustration.
In addition, we assume a weak unfrustrated inter-layer coupling.
It is known that this model leads to a stripe-like antiferromagnetic ordering with wavevectors $(\pi, 0, q_z)$ or $(0, \pi, q_z)$ for sufficiently large $J_2/J_1$. 
Here $q_z = 0, \pi$ depends on the sign of the inter-layer coupling.
In $d = 2$, this collinear order becomes stable for $J_2 / J_1 \gtrsim 0.66$ in the quantum limit ($S=1/2$)~\cite{Richter2010spin} and for $J_2 / J_1 > 0.5$ in the classical limit ($S \to \infty$).
The $J_{1}$-$J_{2}$ Heisenberg model can be derived from the multiband Hubbard model proposed for the iron-based compounds by taking the strong-coupling limit.~\cite{Si2008strong}

A key observation here is that the stripe-like magnetic ordering breaks a discrete $Z_2$ symmetry, associated with two possible bond orderings, in addition to the continuous SU(2) symmetry. 
Therefore, in addition to the spin-wave excitations associated with the broken continuous symmetry, the model with dominant $J_2$ also includes low-energy Ising-like degrees of freedom as 
pointed out  by Chandra \textit{et al}.~\cite{Chandra1990Ising} 
Here, the discrete $Z_2$ symmetry corresponds to a $\pi/2$ rotation of the square lattice.
The bond ordering of the Ising-like degrees of freedom triggers the tetragonal-orthorhombic structural distortion in the presence of a finite spin-lattice coupling.~\cite{Becca2002peierls-like} 
For this reason, the Ising-like ordering is usually interpreted as a structural transition when referring to real compounds.~\cite{Fang2008theory,Xu2008Ising} 
We will denote the transition temperatures for the structural distortion  and the magnetic ordering by $T_{c1} $ and $T_{c2}$, respectively.  
Since the magnetic ordering cannot exist without the Ising ordering, it is clear that $T_{c1}  \geq T_{c2}$~\cite{Fang2008theory,Xu2008Ising} in agreement with experimental observations. 
It is then natural to ask  what are the additional experimental aspects that can be explained qualitatively with a local spin model. 
Finally, the study that we present in this paper is even more relevant for insulating compounds such as
{Li}$_2${V}{O}{Si}{O}$_4$ and {Li}$_2${V}{O}{Ge}{O}$_4$.~\cite{Melzi2000LiVoSi(Ge)O4} 
These compounds are believed to be well described by the $J_1$-$J_2$ model, and an experimental signature of the Ising-like structural transition just above the magnetic transition 
has been suggested for {Li}$_2${V}{O}{Si}{O}$_4$.~\cite{Melzi2000LiVoSi(Ge)O4}

Previous theoretical treatments of the dimensional crossover in the $J_{1}$-$J_{2}$ Heisenberg model at $T>0$ relied on approximate methods such as large-$N$ expansions,~\cite{Fang2008theory} a random-phase approximation (RPA) (or a layer mean-field theory),~\cite{Xu2008Ising} or a phenomenological Landau mean-field theory.~\cite{Cano2010interplay} 
Although all of these treatments agree on that the model has an Ising ordered phase with unbroken spin symmetry in a certain quasi-2D region, the precise form of the phase diagram as a function of the inter-layer coupling is still unknown. 
Moreover, the results obtained by these approximations have several contradicting points relative to the detailed structure of the phase diagram, 
especially when the inter-layer coupling becomes stronger and the two transition temperatures become closer to each other. 
The main motivation of the present study is to resolve these contradictions by applying a controlled numerical method.
We introduce an unfrustrated classical effective model suitable for the $J_{1}$-$J_{2}$ Heisenberg model with dominant $J_2$.
By using a classical Monte Carlo method, we show that the interplay between the Ising and magnetic degrees of freedom leads to a first-order transition when the magnitude of the inter-layer coupling is sufficiently large.
We also provide a renormalization group (RG) argument supporting this observation.
Finally, we numerically identify the Ising ordered phase in the quasi-2D region and present the corresponding phase diagram.
The subtle region where the single transition splits into two transitions is also discussed and compared against recent measurements of two very close transitions in {Ba}{Fe}$_2${As}$_2$.~\cite{Rotundu2011First,Kim2011Character}

\section{%
  \label{sec:model}
  Model
}
We start by considering an unfrustrated classical model that is an effective Hamiltonian for describing the physics of the $J_{1}$-$J_{2}$ Heisenberg model with dominant $J_2$ on a quasi-2D system of weakly coupled square-lattice layers.
The Hamiltonian associated with this so-called Ising-O(3) model~\cite{Yosefin1985phase,Xu2008Ising} is 
\begin{align}
  H = -\sum_{\left\langle i,j\right\rangle}J_{{i}{j}} 
  \left(1 + \sigma_i \sigma_j\right) {\bf S}_i \cdot {\bf S}_j,
  \label{eq:ising-O3}
\end{align}
where $\sigma_i$ and ${\bf S}_i$ denote the classical Ising and O(3) spins, respectively.
The spatially anisotropic coupling constant is:
\begin{align}
  J_{{i}{j}} = 
  \begin{cases}
    J & \text{if $\left(i,j\right)$ are on the same layer,} \\
    J_{z} & \text{if $\left(i,j\right)$ are on the nearest neighbor layers,}
  \end{cases}    
\end{align}
where $0 \le J_{z} \lesssim J$.
In the following we use $J \equiv 1$ as the unit of energy unless otherwise specified.

As we mentioned above, when $J_2 \gtrsim J_{1}/2$, the $J_{1}$-$J_{2}$ Heisenberg model shows magnetic orderings with 
wavevectors $(\pi, 0, q_z)$ or $(0, \pi, q_z)$. 
The stripe-like ordering breaks the lattice rotational symmetry as well as the O(3) spin symmetry.
The ordered state consists of two interpenetrating $\sqrt{2}\times\sqrt{2}$ sublattices, each of which shows a simple {N}{\'e}{e}{l} order, and the inter-sublattice coupling exactly vanishes in the absence of thermal and quantum fluctuations.
Fluctuations stabilize the stripe-like order via the generation of a biquadratic coupling between the two sublattice order parameters favoring the collinear (stripe-like) spin configuration.~\cite{Chandra1990Ising}
This is a clear example of order by disorder.~\cite{Villain1980order,Henley1989ordering,Chandra1990Ising,Kamiya2009finite} 
The  effective Hamiltonian that describes the transition to this broken symmetry state in the long-wavelength limit is:~\cite{Xu2008Ising}
\begin{multline}
  H_{\text{eff}} = \int \mathrm{d}^{d}x\Biggl[~
    \sum_{a = A, B}
    \left(
    \frac{1}{2}
    \left|\nabla{\bm{\phi}}_{a}\right|^2
    + r\lvert{\bm{\phi}}_{a}\rvert^2
    + u\lvert{\bm{\phi}}_{a}\rvert^4
    \right)
    \\
    + u_{AB} \lvert{\bm{\phi}}_{A}\rvert^2 \lvert{\bm{\phi}}_{B}\rvert^2    
    + \lambda \left({\bm{\phi}}_{A}\cdot{\bm{\phi}}_{B}\right)^2
  \Biggr].
  \label{eq:GLW}
\end{multline}
Here, ${\bm{\phi}}_{A}$ and ${\bm{\phi}}_{B}$ represent the three-component sublattice magnetic order parameters. 
The first three terms describe intra-sublattice fluctuations, while the other two quartic terms describe the inter-sublattice couplings allowed by symmetry.
$u$ and $u_{AB}$ are positive, whereas $\lambda$ is negative.
The negative $\lambda$ forces ${\bm{\phi}}_{A}$ and ${\bm{\phi}}_{B}$ to be collinear and
$\langle {\bm{\phi}}_{A} \cdot {\bm{\phi}}_{B} \rangle$ becomes the Ising-like order parameter that decides whether the stripe-like spin configuration is  ``vertical'', as in Fig.~\ref{fig:shematic-pic}, or ``horizontal.''
The model can in principle retain the Ising ordered phase while the magnetic ordering is only of short range.~\cite{Chandra1990Ising,Weber2003ising}
In this phase, we have $\langle {\bm{\phi}}_{A}\rangle = \langle {\bm{\phi}}_{B} \rangle = 0$ but $\langle {\bm{\phi}}_{A} \cdot {\bm{\phi}}_{B}\rangle \ne 0$.~\cite{Chandra1990Ising}

The Ising-O(3) Hamiltonian is an effective model for the $J_1$-$J_2$ Heisenberg model in the sense that both Hamiltonians are described by the same effective theory in the long-wavelength limit if $J_2/J_1$ is sufficiently large. 
The corresponding derivation requires introducing auxiliary O(3) fields $\bm{\phi}_A \sim {\bf S}$ and $\bm{\phi}_B \sim \sigma{\bf S}$, as is described in Appendix~\ref{app:effective hamiltonian}.
The Ising variables,  $\langle\sigma_{i}\rangle$, of the Ising-O(3) model correspond to $\left\langle {\bm{\phi}}_{A} \cdot {\bm{\phi}}_{B} \right\rangle$, while $\langle{\bf S}_i\rangle$ describes the local magnetic ordering.
More heuristically, we can first write down a two-sublattice classical spin model of the form~\cite{Kamiya2010crossover}
\begin{multline}
  H_{\text{coupled}} = -\sum_{\left\langle i,j\right\rangle}J_{{i}{j}} \
  \left(
  {\bf S}_{A,i} \cdot {\bf S}_{A,j}
  + {\bf S}_{B,i} \cdot {\bf S}_{B,j}
  \right)
  \\
  + \sum_{i}
  \lambda \left({\bf S}_{A,i} \cdot {\bf S}_{B,i}\right)^2,
  \label{eq:two coupled model}
\end{multline}
and take the strong biquadratic coupling limit $\lambda \to -\infty$.
Based on symmetry arguments, it is clear that the long-wavelength limit of $H_{\text{coupled}}$ is also described by $H_{\text{eff}}$.  
In the limit $\lambda \to -\infty$, one can write ${\bf S}_{A, i} \equiv {\bf S}_i$ and ${\bf S}_{B, i} = \sigma_{i} {\bf S}_i$ and this procedure yields the Ising-O(3) model.
The Ising-XY model, in which ${\bf S}$ represents a classical XY spin, was studied extensively as an effective model of the fully frustrated XY Hamiltonian,~\cite{Granato1991phase,Lee1991nonuniversal,Nightingale1995conformal}
and an arbitrary coefficient was sometimes included in front of $\sigma_i \sigma_j$ in Eq.~\eqref{eq:ising-O3}.
This prefactor must be equal to unity in the present case because the original $J_1$-$J_2$ model is 
invariant under exchange of the two sublattices $A \leftrightarrows B$.

The factor $J_{{i}{j}} \left(1 + \sigma_i \sigma_j\right)$ in Eq.~\eqref{eq:ising-O3} can be viewed as the effective coupling of the nearest-neighbor classical O(3) spins ${\bf S}_i$ and ${\bf S}_j$.
Because it is non-negative, ferromagnetic alignment of the O(3) spins is always favored. 
This in turn implies that the mean-field coupling, $J_{{i}{j}} \left\langle{\bf S}_i \cdot {\bf S}_j\right\rangle$, between the Ising variables is also ferromagnetic.
The form of $J_{{i}{j}} \left(1 + \sigma_i \sigma_j\right)$ also implies the following restriction on the ordering of the O(3) spins.
Let us consider a case in which the Ising spins are disordered and we divide the system into ferromagnetic clusters of the Ising variables.
We will now quench the Ising spin configuration and perform a partial trace on the O(3) spins.
We see that an O(3) spin of a given cluster cannot correlate with O(3) spins in different clusters because $J_{{i}{j}} \left(1 + \sigma_i \sigma_j\right)$ vanishes at the boundary between the clusters, i.e., on bonds between anti-parallel Ising spins.
Consequently, the O(3) spins cannot order without  Ising ordering, because a percolating 
Ising cluster is required to have O(3) ordering. In this way, we arrive again at the general inequality $T_{c1} \ge T_{c2}$.

In the 2D limit, $J_z = 0$, we expect a finite-temperature Ising transition, while the O(3) spins must remain disordered at any finite 
temperature because of their non-invariance under a continuous symmetry.~\cite{Mermin1966absence}
The Ising transition was confirmed in the 2D classical $J_{1}$-$J_{2}$ Heisenberg model on the square lattice by a Monte Carlo simulation.~\cite{Weber2003ising}
In $d = 2$, we expect that the behavior of the O(3) spins far below the Ising transition should be very close to that of the square-lattice Heisenberg model. 
A finite value of $J_z$ induces a dimensional crossover: 
the Ising transition is shifted to a higher temperature, and the O(3) spins also become ordered at a low enough value of $T$.
A simple RPA argument predicts that $T_{c1} - T_{c1}^\text{2D} \sim \left(J_{z}\right)^{1/\gamma}$, where $\gamma$ is the 2D Ising exponent and $T_{c2} \sim -4\pi/\ln J_{z}$.~\cite{Janke1990crossover,Scalapino1975generalized,Schulz1996dynamics,Yasuda2005neel,Xu2008Ising}
The qualitative difference originates in the different critical behaviors in the 2D limit. 
The power-law correlations of the Ising variables at $T=T_{c1}$ are qualitatively different from the essential singularity at $T = 0$ for the 2D O(3) model.

What should we expect well inside the 3D regime that occurs for large enough $J_z$?
The RPA argument, which describes different order parameters independently, is inappropriate in this regime because the interplay of the Ising and O(3) variables is expected to be strong.
A previous large-$N$ treatment indicates that the two transitions never merge and remain of second order.~\cite{Fang2008theory}
A phenomenological Landau mean-field theory predicts that the O(3) transition should become of first order before merging with the second-order Ising transition.~\cite{Cano2010interplay} 
In addition, this theory predicts a single first-order transition for intermediate values of $J_z$ and a single second-order transition for larger values.

On the other hand, an RG analysis suggests that the merged transition will always be of first order. 
The reason is that a one-loop epsilon expansion applied to a generalization of $H_{\text{eff}}$ in a different context (amorphous magnets) leads to no stable fixed points,~\cite{Aharony1975critical} implying the absence of the scale invariance that is characteristic of second-order transitions (see Appendix~\ref{app:review of RG} for details). 
This RG result contradicts  the above-mentioned Landau theory that allows for a single-second order phase transition when the inter-layer coupling $J_z$ is large enough.~\cite{Cano2010interplay}

\section{%
  \label{sec:method}
  Method
}
We simulated the Ising-O(3) model on the square lattice ($J_z = 0$) and the quasi-2D anisotropic cubic lattice ($0 < J_z < 1$) using the Monte Carlo method.
We employ a cluster Monte Carlo method in which the clusters of Ising and O(3) spins are updated alternatively.
Updates of the O(3) spins take place for fixed  Ising variables 
based on the Wolff algorithm~\cite{Wolff1989collective} and the quenched coupling constant $J_{{i}{j}} \left(1 + \sigma_i \sigma_j\right)$.
The same idea is used for updating the Ising spins, with $J_{{i}{j}} {\bf S}_i \cdot {\bf S}_j$ playing the role of the effective coupling.
The main difference  is that $J_{{i}{j}} {\bf S}_i \cdot {\bf S}_j$ can be negative even though its average is positive.
This fluctuation in the sign of the effective coupling can dynamically induce an effective frustration that makes the cluster updates less efficient.
In practice, it turns out that the fluctuating sign effect does not matter 
for nearly spatially isotropic systems in $d = 2$ and $3$, but it does matter when $J_z$ is small.
To remedy this problem, we incorporate the cluster update with local and semi-global updates of Ising spins. 
In semi-global updates, we allow the clusters to expand only within a given layer by employing the Swendsen-Wang-type multicluster scheme.
The Boltzmann weight factor related to the inter-layer couplings is absorbed in the cluster-flip probability to satisfy detailed balance.
Relatively small intra-layer clusters are expected to be flipped in a collective way by the application of this empirical method.

\section{%
  \label{sec:results}
  Results
}
\begin{figure}[!t]
  \vspace*{-0.40cm}
  \begin{center}
    \includegraphics[angle=270,width=20.5pc, bb=122 77 542 683, clip]{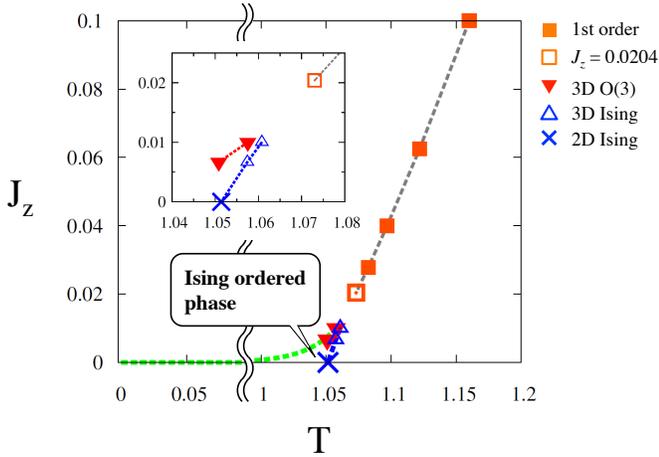}
    \caption{
      (Color online)
      Phase diagram of the quasi-2D Ising-O(3) model.
      The Ising ordered phase ($\langle{\bf S}\rangle = 0$, $\langle\sigma\rangle\ne 0$) appears in the quasi-2D region $J_{z} \lesssim 0.01$.
      For large enough inter-layer coupling, a single first-order transition separates the paramagnetic state ($\langle{\bf S}\rangle = 0$, $\langle\sigma\rangle = 0$) from the lowest-$T$ phase that simultaneously breaks the $Z_2$ and O(3) symmetries ($\langle{\bf S}\rangle \ne 0$, $\langle\sigma\rangle\ne 0$).
      The detailed structure for $0.01 \lesssim J_{z} \lesssim 0.0204$ remains to be clarified; see the text.
      The line representing a phase boundary between the lowest-$T$ phase and the Ising ordered phase is a schematic one. 
      The other lines are guides to the eye.
      \label{fig:phase-diagram}
    }
  \end{center}
\end{figure}

We now show the results of our Monte Carlo simulation.
The Ising-O(3) model exhibits
(i) a second-order transition in the Ising universality class for $J_z = 0$,
(ii) a single first-order transition when $J_z$ is sufficiently large, and
(iii) two second-order transitions in a moderate quasi-2D region ($J_{z} \lesssim 0.01$).
The corresponding  phase diagram is presented in Fig.~\ref{fig:phase-diagram}.
We emphasize that the two transitions merge into a single first-order transition that simultaneously breaks the $Z_2$ and $\mathrm{O}(3)$ symmetries, in agreement with the RG treatment.
As for the split transitions in the quasi-2D region, our numerical results suggest that the transitions are in the 3D Ising and the 3D O(3) universality classes.
There is still some level of uncertainty  relative to the merging of the Ising and O(3) transitions around $0.01 \lesssim J_{z} \lesssim 0.0204$. 
However, our results show some subtle features that agree with the Landau mean-field theory in this region.~\cite{Cano2010interplay}

\subsection{%
  \label{subsec:results:2D}
  Ising transition of the 2D model
}
\begin{figure}[!t]
  \vspace*{-0.40cm}
  \begin{center}
    \includegraphics[width=18pc, bb=70 110 402 542, clip]{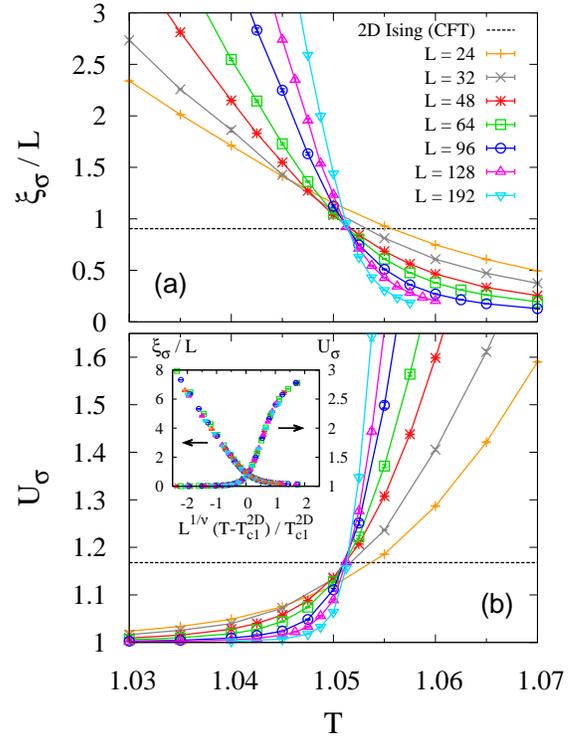}
    \caption{
      (Color online)
      (a) $\xi_\sigma/L$ and 
      (b) $U_{\sigma}$ of the 2D Ising-O(3) model.
      The lines are guides to the eye.
      The horizontal lines marked as ``2D Ising (CFT)'' indicate the universal critical values of the square-lattice Ising model 
      given in Ref.~\onlinecite{Salas2000universal}.
      The inset shows their FSS plots for $L \ge 48$, where $\nu = 1$ and $T_{c1}^\text{2D} = 1.0514(3)$.
      \label{fig:2d-data-ising-U-and-xi}
    }
  \end{center}
\end{figure}
We begin with the 2D Ising-O(3) model.
The maximum lattice size that we studied is $L = 192$.
Due to the effective ferromagnetic coupling between Ising variables generated by the nearest-neighbor O(3) spins, the 2D system undergoes a finite-$T$ transition where only the Ising variables are critical while the O(3) spins remain disordered.
This phase transition can be analyzed very efficiently by introducing the following dimensionless scaling parameters:
the Ising Binder parameter defined by $U_{\sigma} = \langle \sigma^4\rangle / \langle \sigma^2\rangle ^2$, with $\sigma = L^{-2}\sum_i\sigma_i$,
and the second moment correlation length~\cite{Cooper1982solving} of the Ising variables in units of $L$, $\xi_\sigma / L$, with 
\begin{align}
  \xi_\sigma = \sqrt{\frac{\left\langle{\sigma^2}\right\rangle/\left\langle{\sigma_{\bf q}^2}\right\rangle - 1}{4\sin^2\left(\pi/L\right)}}.
\end{align}
Here $\sigma_{\bf q}$ is the Fourier mode at the lowest nonzero momentum, ${\bf q} = \left(2\pi/L,0\right)$ or $\left(0,2\pi/L\right)$, for a given lattice.
\begin{figure}[!b]
  \vspace*{-0.40cm}
  \begin{center}
    \includegraphics[width=18pc, bb=70 300 402 542, clip]{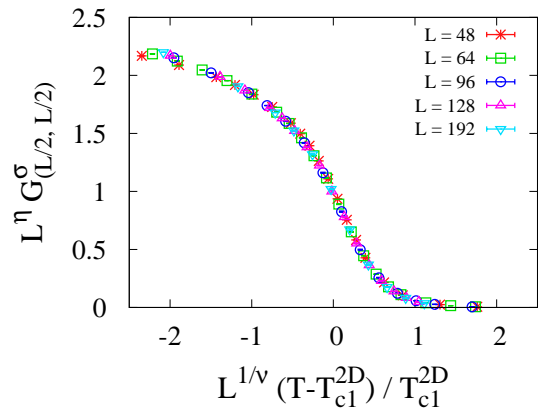}
    \caption{
      (Color online)
      FSS of the Ising correlation function $G_{\left(L/2, L/2\right)}^{\sigma}$ 
      at the largest distance in a given 2D system defined by Eq.~\eqref{eq:Ising-correlation-function-2D-at-the-largest-distance}, 
      where $\eta = 0.25$, $\nu = 1$, and $T_{c1}^\text{2D} = 1.0514(3)$.
      \label{fig:2d-data-ising-Gs}
    }
  \end{center}
\end{figure}
These variables asymptotically cross each other at the Ising transition point as a function of the system size.
As shown in Fig.~\ref{fig:2d-data-ising-U-and-xi}, the crossing point found for the larger lattices leads to a value of $T_{c1}^\text{2D} = 1.0514(3)$. 
It is also known that the values of these scaling parameters should be universal at their crossing points. 
In agreement with this expectation, our results are consistent with the universal values for the 2D Ising model, which are obtained by using the conformal field theory (see Fig.~\ref{fig:2d-data-ising-U-and-xi}).~\cite{Salas2000universal}
Our final confirmation of the 2D Ising universality class is a finite-size scaling (FSS) plot of the dimensionless quantities shown in the inset of Fig.~\ref{fig:2d-data-ising-U-and-xi}, and another FSS plot of the correlation function for the Ising spins at the most distant sites of a given system size (see Fig.~\ref{fig:2d-data-ising-Gs}),
\begin{align}
  G_{\left(L/2, L/2\right)}^{\sigma} = \left\langle\sigma_i\sigma_j\right\rangle~~\text{with}~~
  {\bf r}_{{i}{j}} = \left(L/2, L/2\right).
  \label{eq:Ising-correlation-function-2D-at-the-largest-distance}
\end{align}
We assume $\eta = 0.25$ and $\nu = 1$ in these FSS plots.
\begin{figure}[!t]
  \vspace*{-0.40cm}
  \begin{center}
    \leavevmode
    \includegraphics[width=18pc, bb=70 290 402 542, clip]{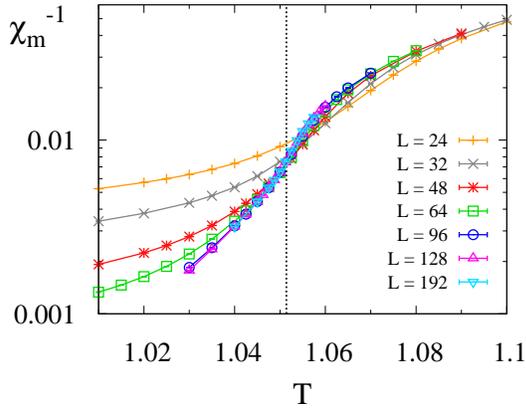}
    \caption{
      (Color online)
      The inverse susceptibility of the O(3) order parameter of the 2D Ising-O(3) model ($J_z = 0$).
      The vertical line indicates the 2D Ising transition temperature $T_{c1}^\text{2D}$.
      \label{fig:2d-data-ising-inv-chi-O3}
    }
  \end{center}
\end{figure}

\subsection{%
  \label{subsec:results:Q2D}
  Separate transitions in the quasi-2D regime
}
We now discuss the quasi-2D regime.
The finite inter-layer coupling makes the Ising order three-dimensional and it also stabilizes the O(3) spin order at $T > 0$.
Of our particular interest, motivated by the separate transitions in the 1111 compounds, is the region where the Ising and the O(3) transitions occur at different temperatures.
The order of magnitude of the inter-layer coupling in such a region can be estimated from a simple RPA argument.~\cite{Janke1990crossover,Scalapino1975generalized,Schulz1996dynamics,Yasuda2005neel,Xu2008Ising}
The previous RPA estimates of $T_{c1}$ and $T_{c2}$ were obtained from the following conditions: $J_{z}\chi_{\sigma}^\text{2D} \left(T_{c1}\right) \sim 1$ and $J_{z}\chi_{m}^\text{2D} \left(T_{c2}\right) \sim 1$ where $\chi_{\sigma}^\text{2D}$ and $\chi_{m}^\text{2D}$ are the susceptibilities of the Ising and O(3) variables for $J_z = 0$.
To obtain a rough estimate of $J_z$ in the regime of separate transitions, 
we define $J_{z}^{\ast}$ as the magnitude of the inter-layer coupling for which the RPA estimate of $T_{c2}$ coincides with $T_{c1}^\text{2D} = 1.0514(3)$, i.e., $(J_{z}^{\ast})^{-1} \equiv \chi_{m}^\text{2D} \left(T_{c1}^\text{2D}\right)$.
By using our numerical estimation of $\chi_{m}^\text{2D}$ shown in Fig.~\ref{fig:2d-data-ising-inv-chi-O3}, we obtain $J_{z}^{\ast} = 0.00765(2)$.
Figure~\ref{fig:2d-data-ising-inv-chi-O3}, in conjunction with the RPA argument, also suggests that $J_{z} \lesssim 0.001$ is required to obtain a separation $T_{c1} - T_{c2}$ of order $0.01J$.

\begin{figure}[!t]
  \vspace*{-0.40cm}
  \begin{center}
    \includegraphics[width=18pc, bb=50 45 410 790, clip]{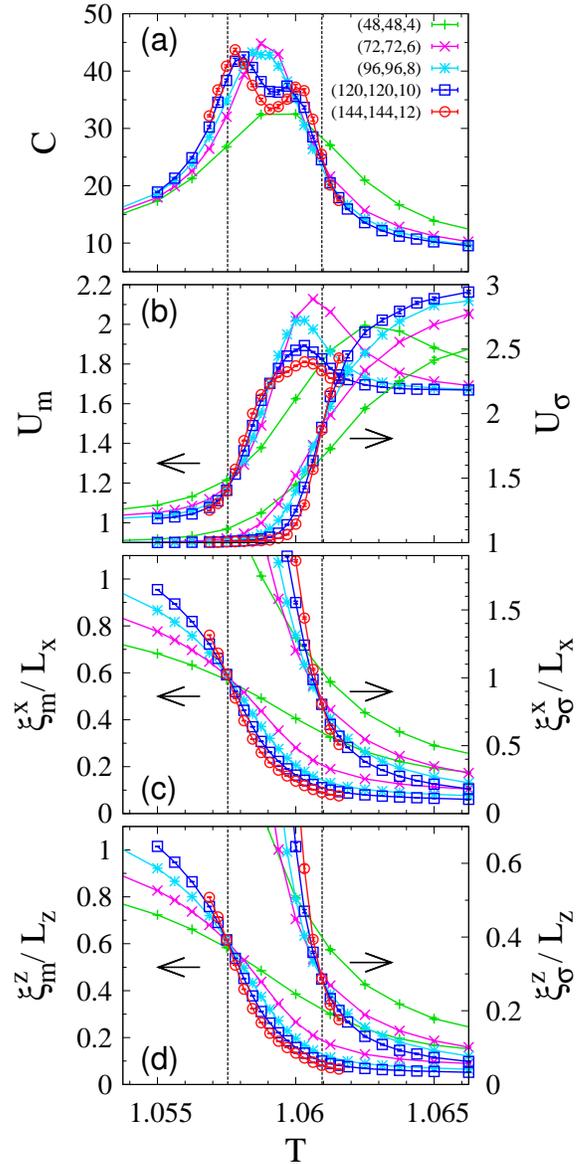}
    \caption{
      (Color online)
      Two separate transitions for $J_z = 0.01$:
      (a) specific heat,
      (b) the Binder parameters,
      (c) the normalized correlation lengths in the $x$ direction, and
      (d) the normalized correlation lengths in the $z$ direction.
      The vertical lines show the estimated critical temperatures
      based on the FSS analysis.
      The lines are guides to the eye.
      \label{fig:q2d-rawdata}
    }
  \end{center}
\end{figure}

Unfortunately, a direct finite-size study for $J_{z} \lesssim 0.001$ is not simple because the highly anisotropic correlations would require  very large system sizes even for simulating a few  layers (see, e.g., Ref.~\onlinecite{Yasuda2005neel}).
For this reason, we concentrate on $J_{z} = 0.01$.
(We also simulated the model for $J_{z} = 0.006667$ and observed essentially the same phenomena.)
Although this value of $J_z$ is slightly larger than $J_{z}^{\ast}$, the two transitions still occur at different temperatures.
To reduce the finite-size effects induced by the spatial anisotropy, the system is set to have a tetragonal shape 
$L_x \times L_y \times L_z$ with periodic boundary conditions ($L_x = L_y \equiv L$).
The aspect ratio $r \equiv L_z / L = 1/12$  was determined in such a way that $G^{\sigma}(L_x/2, 0, 0) \approx G^{\sigma}(0, 0, L_z/2)$ and $G^{m}(L_x/2, 0, 0) \approx G^{m}(0, 0, L_z/2)$, where
\begin{align}
  G^{\sigma}(r_x, r_y, r_z) = \left\langle\sigma_i\sigma_j\right\rangle,~~
  G^{m}(r_x, r_y, r_z) = \left\langle{\bf S}_i\cdot{\bf S}_j\right\rangle
  \label{eq:correlation-functions-3D}
\end{align}
with ${\bf r}_{{i}{j}} = \left(r_x, r_y, r_z\right)$ are the correlation functions of the Ising and the O(3) spins.
We studied a range of system sizes from $L = 48$ to $L = 144$.

\begin{figure}[!t]
  \vspace*{-0.40cm}
  \begin{center}
    \includegraphics[width=18pc, bb=50 50 410 720, clip]{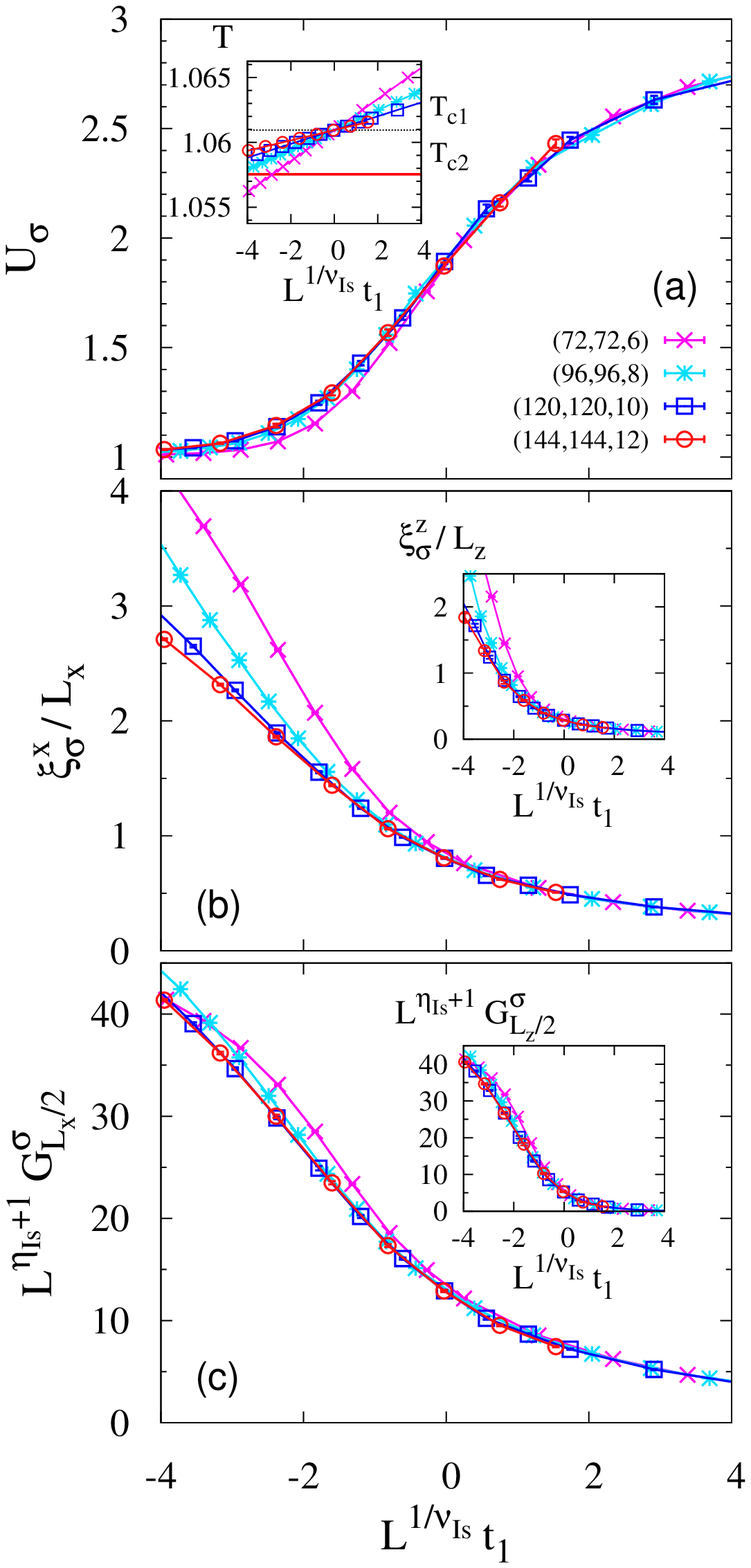}
    \caption{
      (Color online)
      FSS plots for the higher-$T$ Ising transition for $J_z = 0.01$ [$t_{1} \equiv \left(T - T_{c1}\right)/T_{c1}$].
      The upper panel (a) shows the FSS of $U_{\sigma}$. 
      The inset shows the relation to the real temperature $T$; 
      the thick line indicates the lower-$T$ transition close to which a severe finite-size effect appears.
      The middle panel (b) shows the FSS of $\xi_{\sigma}^{x}/L_x$, with the inset
      showing the FSS of $\xi_{\sigma}^{z}/L_z$.
      The lower panel (c) shows the FSS of 
      $G^{\sigma}_{L_x/2} \equiv G^{\sigma}(L_x/2, 0, 0)$ and $G^{\sigma}_{L_z/2} \equiv G^{\sigma}(0, 0, L_z/2)$ (the inset)
      [see Eq.~\eqref{eq:correlation-functions-3D}].
      \label{fig:q2d-fss-Z2}
    }
  \end{center}
\end{figure}
\begin{figure}[!t]
  \vspace*{-0.40cm}
  \begin{center}
    \includegraphics[width=18pc, bb=50 50 410 720, clip]{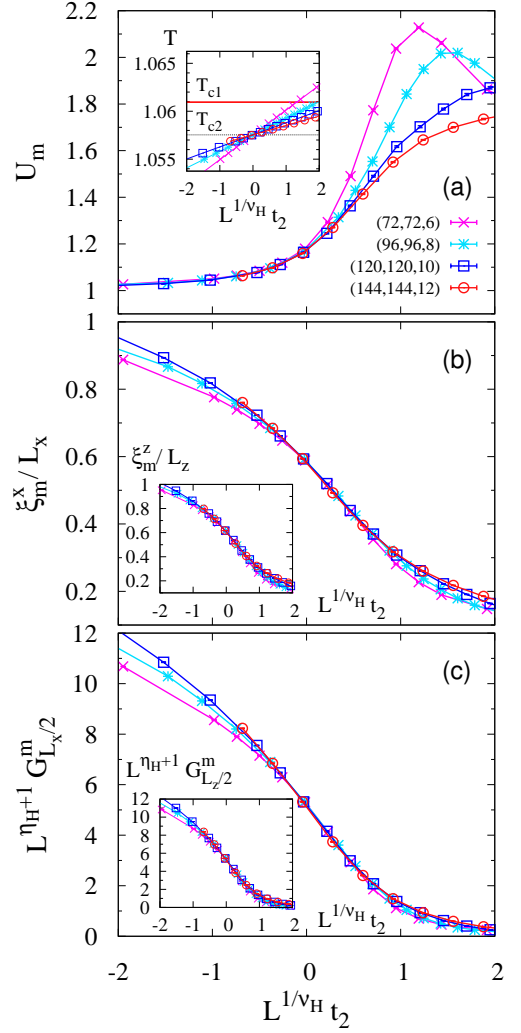}
    \caption{
      (Color online)
      FSS plots for the lower-$T$ O(3) transition for $J_z = 0.01$ [$t_{2} \equiv \left(T - T_{c2}\right)/T_{c2}$].
      The upper panel (a) shows the FSS of $U_m$.
      The inset shows the relation to the real temperature $T$;
      the thick line indicates the higher-$T$ transition close to which a severe finite-size effect appears.
      The middle panel (b) shows the FSS of $\xi_{m}^{x}/L_x$, with the inset
      showing the FSS of $\xi_{m}^{z}/L_z$.
      The lower panel (c) shows the FSS of 
      $G^{m}_{L_x/2} \equiv G^{m}(L_x/2, 0, 0)$ and $G^{m}_{L_z/2} \equiv G^{m}(0, 0, L_z/2)$ (the inset) 
      [see Eq.~\eqref{eq:correlation-functions-3D}].
      \label{fig:q2d-fss-O3}
    }
  \end{center}
\end{figure}
In Fig.~\ref{fig:q2d-rawdata}(a), we show the specific heat per site, $C = N^{-1} \beta^2\bigl(\left\langle H^2 \right\rangle - \left\langle H \right\rangle^2\bigr)$, where $N = {r}{L}^3$ is the number of sites.
The double-peak structure of the specific heat with increasing depth for
larger systems gives a first indication of two separate transitions.
This observation is supported by the behavior of the dimensionless scaling parameters.
Here we use the Binder parameters, $U_{\sigma} = \langle\sigma^4\rangle / \langle\sigma^2\rangle^2$ with $\sigma = N^{-1}\sum_i\sigma_i$ and $U_{m} = \langle{{\bf m}}^4\rangle / \langle {{\bf m}}^2\rangle^2$ with ${{\bf m}} = N^{-1}\sum_i{\bf S}_i$, and the normalized correlation lengths, $\xi_{\sigma}^{\mu}/L^{\mu}$ and $\xi_{m}^{\mu}/L^{\mu}$, of the Ising and the O(3) order parameters along the intra-layer ($\mu = x$) and the inter-layer ($\mu = z$) directions.
$\xi_{\sigma}^{\mu}$ and $\xi_{m}^{\mu}$  are defined by:
\begin{align}
  \xi_{\sigma}^{\mu} &= \sqrt{\frac{\left\langle{\sigma_{{\;}_{\;}}^2}\right\rangle/\left\langle{\sigma_{{\bf q}_{\mu}}^2}\right\rangle - 1}{4\sin^2\left(\pi/L_{\mu}\right)}},
  \\
  \xi_{m}^{\mu} &= \sqrt{\frac{\left\langle{\left\lvert{\bf m}\right\rvert^2}\right\rangle/\left\langle{\left\lvert{\bf m}_{{\bf q}_{\mu}}\right\rvert^2}\right\rangle - 1}{4\sin^2\left(\pi/L_{\mu}\right)}},
\end{align}
where ${\bf q}_{\mu}$ is the lowest nonzero momentum for a given lattice in the $\mu$ direction.
As shown in Figs.~\ref{fig:q2d-rawdata}(b)--\ref{fig:q2d-rawdata}(d), the curves of these quantities exhibit a clear tendency to intersect for larger lattices, and the crossing temperature \rm{varies depending on the} order parameter to which they are related. 
While we find no peak in $U_{\sigma}$, there is a peak structure in $U_{m}$ for small systems, but this is suppressed for larger lattices.
The non-divergent behavior of the Binder parameters \rm{eliminates} the possibility of strongly first-order transitions.~\cite{Vollmayr1991finite}
The FSS plots are presented in Figs.~\ref{fig:q2d-fss-Z2} and \ref{fig:q2d-fss-O3}.
Based on symmetry arguments, we assume that the higher-$T$ (lower-$T$) transition is in the 3D Ising [O(3)] universality class, 
and we use the corresponding critical exponents that are available in the literature:
$\eta_{\mathrm{Is}} = 0.03639(15)$ and $\nu_{\mathrm{Is}} = 0.63012(16)$ for the 3D Ising universality class~\cite{Campostrini200225th-order} and $\eta_{\mathrm{H}} = 0.0375(5)$ and $\nu_{\mathrm{H}} = 0.7112(5)$ for the 3D O(3) universality class.~\cite{Campostrini2002critical}
Although finite-size effects are still severe for the explored system sizes, we can see an asymptotic tendency toward data collapse. 
This observation supports the assumed universality classes, which leads to $T_{c1} = 1.0610(7)$ and $T_{c2} = 1.0575(2)$.
The rather small separation is indeed expected because $J_z = 0.01$ is relatively large in comparison with $J_{z}^{\ast}$.

As for the sizable sub-leading corrections observed in these scaling plots, it appears that they are largely due to the proximity of the two transitions and/or the spatial anisotropy of the correlations.
Naturally, the proximity effect is expected to appear in the low- (high-) temperature side of the scaling plots for the Ising [O(3)] transition.
For instance,  a deviation appears in the scaling plots around the O(3) transition  for $L^{1/\nu_{\mathrm{H}}}\left(T - T_{c2}\right)/T_{c2} \gtrsim 1$  and  $L = 72$ (Fig.~\ref{fig:q2d-fss-O3}). 
This deviation is most likely due to the proximity to the Ising transition.
A similar effect is observed in the FSS plots of the Ising-like transition for $L^{1/\nu_{\mathrm{Is}}}\left(T - T_{c1}\right)/T_{c1} \lesssim -2$ and $L = 72$ (Fig.~\ref{fig:q2d-fss-Z2}).
However, these finite-size effects  disappear rapidly for larger system-sizes.
As for the finite-size effect due to spatial anisotropy, we find that the crossing value of $\xi_{\sigma}^{\mu}/L_{\mu}$ depends on $\mu = x, z$, although the crossing value of $\xi_{m}^{\mu}/L_{\mu}$ is almost independent of $\mu$ [Figs.~\ref{fig:q2d-rawdata}(c) and \ref{fig:q2d-rawdata}(d)].
This observation suggests that the aspect ratio $r = 1/12$ for $J_z = 0.01$ is appropriately tuned to investigate the O(3) transition of this system, but it is not perfectly tuned for investigating the Ising transition. 
The observation of $\xi_{\sigma}^{x}/L_{x} > \xi_{\sigma}^{z}/L_{z}$ implies a shortness of the effective inter-layer coupling of the Ising spins in the simulated finite system (with $r =1/12$) and such an anisotropy effect might produce sub-leading corrections to the scaling behavior near $T = T_{c1}$. 
However, we believe that these corrections will not affect our conclusions significantly.
In particular, our conclusion about the separation of the transitions for $J_{z} = 0.01$ does not change.
The reason is that our estimation of $T_{c2}$ is accurate enough and a modified aspect ratio of $r < 1/12$ will never lower the estimation of $T_{c1}$
 because it has the effect of enhancing the effective inter-layer coupling between Ising variables.

\subsection{%
  \label{subsec:results:3D}
  3D system with a large inter-layer coupling
}
\begin{figure}[!t]
  \vspace*{-0.40cm}
  \begin{center}
    \includegraphics[width=20pc, bb=65 50 445 278, clip]{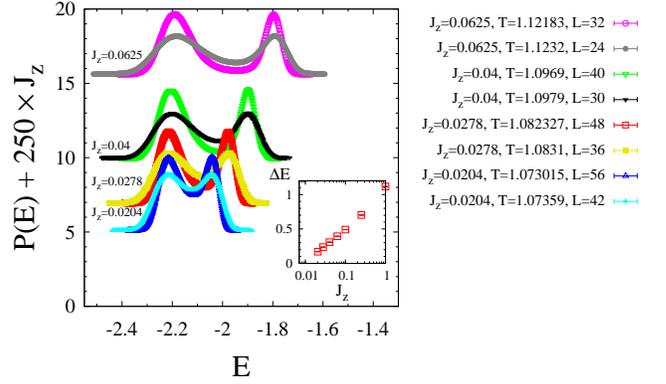}
    \caption{
      (Color online)
      Bimodal internal energy density distribution at the first order phase transition.
      The inset shows the $J_{z}$-dependence of the peak-to-peak distance $\Delta E$ of the distributions.
      \label{fig:3d-1st-order}
    }
  \end{center}
\end{figure}
\begin{figure}[!t]
  \vspace*{-0.20cm}
  \begin{center}
    \includegraphics[width=20.5pc, bb=85 480 765 885, clip]{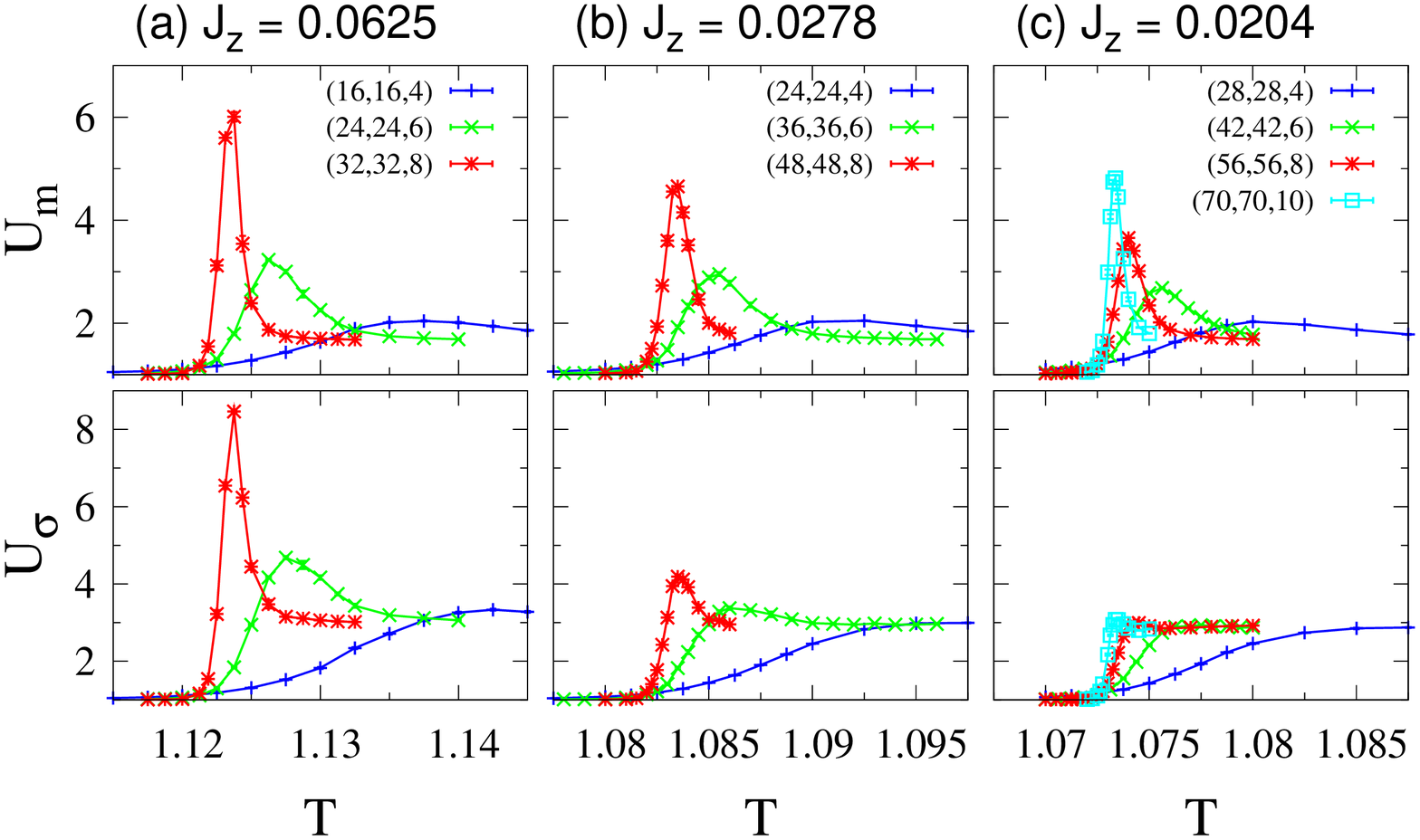}
    \caption{
      (Color online)
      Binder parameters $U_m$ and $U_\sigma$ near the first order
      transition for
      (a) $J_{z} = 0.0625$, 
      (b) $J_{z} = 0.0278$ and
      (c) $J_{z} = 0.0204$.
      \label{fig:1st-order-U}
    }
  \end{center}
\end{figure}
We now discuss the 3D regime, where the inter-layer coupling is sufficiently large.
We note that previous analytical studies suggested different scenarios in this region.
A large-$N$ approximation~\cite{Fang2008theory} predicted a scenario with two separate second-order transitions, whereas a phenomenological Landau mean-field theory predicted a richer structure.~\cite{Cano2010interplay}
The Landau theory also suggested a possibility of a single second-order transition.~\cite{Cano2010interplay}
We simulated the system for $J_{z} = 1$, $0.25$, $0.1$, $0.0625$, $0.04$, $0.0278$, and $0.0204$.
The aspect ratios of the lattices are $r = 1$, $1/2$, $1/3$, $1/4$, $1/5$, $1/6$, and $1/7$, respectively, and were determined to reduce finite-size effects.

In Fig.~\ref{fig:3d-1st-order}, we show the internal energy distribution at a temperature around which the Binder parameters show characteristic features of a phase transition (a shift from the high-$T$ Gaussian value to the trivial low-$T$ value) for several values of $J_{z}$.
The distribution exhibits bimodal structure, which is an unambiguous signature of a first-order transition. 
The peak-to-peak distance, $\Delta E$, is a finite-size estimate of the latent heat. 
As shown in the inset of Fig.~\ref{fig:3d-1st-order}, $\Delta E$  increases monotonically as a function of $J_{z}$ in this region.

It is natural to ask what are the broken symmetries below the first-order transition.
In Figs.~\ref{fig:1st-order-U}(a) and \ref{fig:1st-order-U}(b) we show the temperature dependence of the Binder parameters $U_{\sigma}$ and $U_{m}$ for $J_{z} = 0.0625$ and $0.0278$, respectively.
Both of them show diverging behavior, indicating discontinuity of the corresponding order parameters.~\cite{Vollmayr1991finite}
The same behavior is observed for the $J_z$ values listed above, except for the case $J_{z} = 0.0204$, which will be discussed later.
The peaks are sharper for larger $J_z$ values.
Based on this observation and the monotonically increasing value of $\Delta E$, we conclude that the 3D Ising-O(3) model with a large enough inter-layer coupling undergoes a single first-order transition.
Our conclusion agrees with the RG analysis of $H_{\text{eff}}$,~\cite{Aharony1975critical} while it discards the other previous scenarios, namely the single second-order transition suggested by the Landau mean-field theory~\cite{Cano2010interplay} or the always separate second-order transitions suggested by the previous large-$N$ approximation.~\cite{Fang2008theory}
However, recently another large-$N$ approach yielded a phase diagram showing a single first-order transition when the Ising and the O(3) transitions are merged, in agreement with our results.~\cite{FernandesArXivPreemptive}
On the other hand, the failure of the Landau mean-field theory at this point is not surprising because the system is below the upper critical dimension $d = 4$.

\begin{figure}[!t]
  \vspace*{-0.40cm}
  \begin{center}
    \includegraphics[angle=90, width=18pc, bb=156 187 410 668, clip]{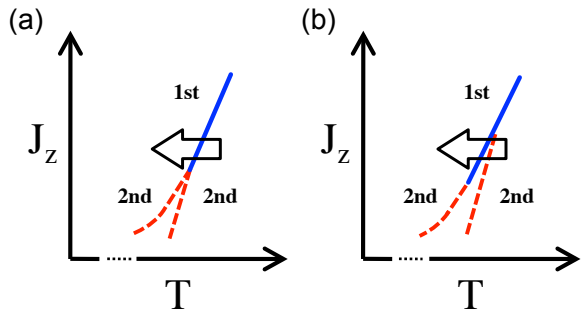}
    \caption{
      (Color online)
      Two possible scenarios that may explain our observations in the intervening region are shown.
      (a) The two transitions continue to be of second order until they collapse into the direct first-order transition.
      (b) Only the lower-$T$ O(3) transition becomes of first order before the transitions merge.
      The arrows schematically indicate the possible ways in which the system is cooled down for $0.01 \lesssim J_z \lesssim 0.0204$.
      \label{fig:scenarios}
    }
  \end{center}
\end{figure}

Finally, we briefly discuss the region where the two transitions merge. 
As we mentioned before, there is some level of uncertainty in this  intervening region.
By comparing Fig.~\ref{fig:1st-order-U}(a) and \ref{fig:1st-order-U}(b), we notice that there is a monotonic tendency in the peak structure. 
As $J_z$ decreases, the peak of $U_{\sigma}$ is drastically suppressed as compared to that of $U_{m}$.
Indeed, as shown in Fig.~\ref{fig:1st-order-U}(c), $U_{\sigma}$ for $J_{z} = 0.0204$ does not exhibit an evident diverging behavior for the explored system sizes, while $U_{m}$ clearly does.
One possible explanation is that finite-size effects smear the discontinuity of the Ising order parameter because the Ising correlation length is larger than $L$ (not shown). 
Thus, based on our numerical results, we cannot discard a scenario in which the direct first-order transition to the lowest-$T$ ordered phase ends up at a critical point where it starts splitting into two second-order transitions [see Fig.~\ref{fig:scenarios}(a)].
However, our RG analysis indicates that such a critical end point would not be stable.
The most plausible scenario corresponds to the existence of a finite region where the system in the lowest-$T$ phase first recovers the O(3) symmetry via a first-order transition, while the $Z_2$ symmetry is recovered at a higher temperature via a second-order Ising-like transition [see Fig.~\ref{fig:scenarios}(b)].
This implies that $U_{m}$ should diverge around the first-order transition, whereas $U_\sigma$ should reach a trivial and finite low-$T$ value below $T = T_{c1}$ for large enough system sizes.
Although we still do not have enough evidence to confirm this scenario, our results suggest that it may occur near $J_z = 0.0204$.

\section{%
  \label{sec:summary}
  Summary
}
In summary, we have studied the Ising-O(3) model on quasi-2D lattices. 
This is an effective model of the $J_{1}$-$J_{2}$ Heisenberg model in which $J_2$ is dominant.
By solving this effective Hamiltonian, we identified the region where the O(3)-symmetric Ising ordered phase exists.
For sufficiently large inter-layer coupling, we found that a single first-order transition occurs between the paramagnetic phase and the lowest-$T$ ordered phase, in agreement with the previous RG treatment on $H_{\text{eff}}$.~\cite{Aharony1975critical}
Although the question of how this first-order transition splits into two transitions remains as an open problem, the scenario shown in Fig.~\ref{fig:scenarios}(b) provides the most reasonable explanation of our numerical results.

Our results provide a qualitative explanation for the sequence of transitions observed in ferropnictides as a function of the ratio between the inter-layer and intra-layer exchange couplings.
According to these results, the separate structural and SDW transitions observed in the quasi-2D 1111 compounds~\cite{Cruz2008EXP:magnetic,McGuire2008phase} are caused by the fragility of the continuous SDW order against fluctuations, which makes it more sensitive to the magnitude of the inter-layer coupling.
The suggested structural transition in
{Li}$_2${V}{O}{Si}{O}$_4$~\cite{Melzi2000LiVoSi(Ge)O4} is also naturally explained by the same mechanism.
The first-order nature of the simultaneous structural and SDW transition observed in most of the more 3D 122 parent compounds~\cite{Krellner2008magnetic,Goldman2008Lattice,Yan2008Structural} is also consistent with our results.
This is related to the absence of the stable RG fixed point.~\cite{Aharony1975critical}
Remarkably, a recent measurement also found that a sequence of transitions that is entirely consistent with the scenario presented in Fig.~\ref{fig:scenarios}(b) takes place in {Ba}{Fe}$_2${As}$_2$.~\cite{Rotundu2011First,Kim2011Character}
Therefore, despite the oversimplified nature of our local-moment model for a microscopic description of the metallic ferropnictides, we have reproduced their qualitative phase diagram (Fig.~\ref{fig:phase-diagram}). 
Our results thus indicate that the $J_{1}$-$J_{2}$ Heisenberg model and the related Ising-O(3) model are good starting points for describing the universal properties of these compounds.

\begin{acknowledgments}
We acknowledge M. Oshikawa, H. Tsunetsugu, S. Miyashita, M. Takigawa, T. Sakakibara, and R. M. Fernandes for valuable discussions and comments.
The computation in the present work was executed at the Supercomputer Center, Institute for Solid State Physics, University of Tokyo.
This work is financially supported by
GCOE for Phys. Sci. Frontier, MEXT, Japan,
the MEXT Grants-in-Aid for Scientific Research (B) (22340111), 
the MEXT Grants-in-Aid for Scientific Research on Priority Areas
``Novel States of Matter Induced by Frustration'' (19052004), 
and the Next Generation Supercomputing Project, Nanoscience Program, MEXT, Japan.
Work at Los Alamos National Laboratory was performed under the auspices of the U.S.\ DOE, Contract No.~DE-AC52-06NA25396, through the LDRD program.
\end{acknowledgments}

\appendix
\section{%
  \label{app:effective hamiltonian}
  EFFECTIVE HAMILTONIAN FOR THE ISING-O($N$) MODEL
}
Let us consider a generalization of the Ising-O(3) model replacing the three-component spin by the O($N$) spin:
$
H = -\sum_{\langle {i}{j} \rangle} 
J_{{i}{j}}
\left(1+\sigma_i\sigma_j\right){\bf S}_{i} \cdot {\bf S}_{j}
$,
where $\sigma_i$ and ${\bf S}_i$ are the Ising and the O($N$) spins, respectively.
In the following, we derive an effective Hamiltonian for this model.
By using a Gaussian transformation, we introduce the auxiliary $N$-component vector fields 
$\bm{\phi}_{A} \sim {\bf S}$ 
and 
$\bm{\phi}_{B} \sim \sigma{\bf S}$.
By introducing $K_{{i}{j}} = \beta J_{{i}{j}}$, we obtain:
\begin{align}
  &
  e^{-\beta H} =
  \exp\left(\frac{1}{2}~ {\bf S}_{i} \cdot K_{{i}{j}} {\bf S}_{j}\right)~
  \exp\left(\frac{1}{2}~ \sigma_i{\bf S}_{i} \cdot K_{{i}{j}} \sigma_j{\bf S}_{j}\right)
  \notag
  \displaybreak[0]
  \\
  &\propto
  \int D[\bm{\phi}_A] D[\bm{\phi}_B]
  \exp\left(
  -\frac{1}{2} \bm{\phi}_{A,i} \cdot K^{-1}_{{i}{j}} \bm{\phi}_{A,j}
  + \bm{\phi}_{A,i} \cdot {\bf S}_{i}
  \right)
  \notag
  \\
  &\hspace{10pt}
  \times\exp\left(
  -\frac{1}{2} \bm{\phi}_{B,i} \cdot K^{-1}_{{i}{j}} \bm{\phi}_{B,j}
  + \bm{\phi}_{B,i} \cdot \sigma_i{\bf S}_{i}
  \right)
  \notag
  \displaybreak[0]
  \\
  &= 
  \int D[\bm{\phi}_A] D[\bm{\phi}_B]
  \exp\left(
  -\frac{1}{2} \sum_{a = A, B}\bm{\phi}_{a,i} \cdot K^{-1}_{{i}{j}} \bm{\phi}_{a,j}
  \right)
  \notag
  \\
  &\hspace{10pt}
  \times
  \exp\left[
    \left(\bm{\phi}_{A,i} + \sigma\bm{\phi}_{B,i}\right) \cdot {\bf S}_{i}
  \right]
  \label{eq:app:1}
\end{align}
where the summation rule for duplicate indices is assumed.
Since we have decoupled the spins on different sites, we can trace them out on each site:
\begin{align}
  Z &\propto
  \int D[\bm{\phi}_A] D[\bm{\phi}_B]
  \exp\left(
  -\frac{1}{2} \sum_{a = A, B}\bm{\phi}_{a,i} \cdot K^{-1}_{{i}{j}} \bm{\phi}_{a,j}
  \right)
  \notag\\
  &\hspace{64pt}
  \times\prod_{i}
  \mathrm{Tr}_{{\bf S}, \sigma}
  \exp\left[
    \left(\bm{\phi}_{A,i} + \sigma\bm{\phi}_{B,i}\right) \cdot {\bf S}
  \right]
  \notag
  \displaybreak[0]
  \\
  &=
  \int D[\bm{\phi}_A] D[\bm{\phi}_B]
  \exp\left(
  -\frac{1}{2} \sum_{a = A, B}\bm{\phi}_{a,i} \cdot K^{-1}_{{i}{j}} \bm{\phi}_{a,j}
  \right)
  \notag\\
  &\hspace{32pt}
  \times\exp\left\{
  \sum_i
  \ln\left[
    \sum_{\sigma = \pm 1}
    G_N(\bm{\phi}_{A,i} + \sigma \bm{\phi}_{B,i})
    \right]
  \right\}.
  \label{eq:app:2}
\end{align}
Here, 
\begin{align}
  G_{N}(\bm{j}) 
  & = 
  \mathrm{Tr}_{{\bf S}}
  \exp \left( \bm{j} \cdot {\bf S}\right)
  \displaybreak[0]
  \notag\\
  &= 
  \frac{
    \int d^{N}S\, \delta\left({{\bf S}^2 - 1}\right)
    \exp \left( \bm{j} \cdot {\bf S}\right)
  }{
    \int d^{N}S\, \delta\left({{\bf S}^2 - 1}\right)
  }
  \displaybreak[0]
  \notag\\
  &= \sum_{n = 0}^{\infty} c_{n} \left( \bm{j} \cdot \bm{j}\right)^n,~~
  c_{n} = \frac{\Gamma(N/2)}{2^{2n}~n!~\Gamma(N/2+n)}
  \label{eq:app:3}
\end{align}
is a single-site generating function. 
The following terms appear in the expansion of $\sum_{\sigma = \pm 1}G_N(\bm{\phi}_{A,i} + \sigma\bm{\phi}_{B,i})$:
\begin{align}
  &
  c_2 \left(
  \left\lvert 
  \bm{\phi}_{A,i} + \bm{\phi}_{B,i}
  \right\rvert^2
  +
  \left\lvert 
  \bm{\phi}_{A,i} - \bm{\phi}_{B,i}
  \right\rvert^2
  \right)
  \displaybreak[0]
  \notag\\
  &\hspace{10pt}=
  2c_2 \left(
  \left\lvert 
  \bm{\phi}_{A,i}
  \right\rvert^2
  +
  \left\lvert 
  \bm{\phi}_{B,i}
  \right\rvert^2
  \right)
  \notag\\
  &\hspace{10pt}\equiv 2c_2 \Phi_2,
  \displaybreak[0]
  \\[+6pt]
  &
  c_4 \left(
  \left\lvert 
  \bm{\phi}_{A,i} + \bm{\phi}_{B,i}
  \right\rvert^4
  +
  \left\lvert 
  \bm{\phi}_{A,i} - \bm{\phi}_{B,i}
  \right\rvert^4
  \right)
  \displaybreak[0]
  \notag\\
  &\hspace{10pt}=
  2c_4 \left[
    \left(
    \left\lvert 
    \bm{\phi}_{A,i}
    \right\rvert^2
    +
    \left\lvert 
    \bm{\phi}_{B,i}
    \right\rvert^2
    \right)^2
    + 4 
    \left(
    \bm{\phi}_{A,i}\cdot\bm{\phi}_{B,i}
    \right)^2
    \right]
  \notag\\
  &\hspace{10pt}\equiv 2c_4 \Phi_4.
\end{align}
By truncating at fourth order in powers of the fields $\bm{\phi}_{A}$ and $\bm{\phi}_{B}$, we obtain:
\begin{align}
  &
  \ln\left[
    \sum_{\sigma = \pm 1}
    G_N(\bm{\phi}_{A,i} + \sigma \bm{\phi}_{B,i})
    \right]
  \notag\\[+2pt]
  &\hspace{10pt}=
  \ln\left(
  c_1 
  + 2c_2 \Phi_2
  + 2c_4 \Phi_4
  + \dots
  \right)
  \displaybreak[0]
  \notag\\
  &\hspace{10pt}\simeq
  \ln c_1
  + \frac{2c_2}{c_1} \Phi_2
  + \frac{2c_4}{c_1} \Phi_4
  - \frac{1}{2}\left(\frac{2c_2}{c_1}\right)^2 (\Phi_2)^2
  + \dots
\end{align}
From this expression we obtain
\begin{align}
  Z &= \int D[\bm{\phi}_A] D[\bm{\phi}_B]
  \exp\left(-S[\bm{\phi}_A, \bm{\phi}_B]\right),
\end{align}
with
\begin{align}
  &
  S[\bm{\phi}_A, \bm{\phi}_B]
  \notag\\[+2pt]
  &=
  \frac{1}{2} \sum_{a=A,B}\bm{\phi}_{a,i} \cdot K^{-1}_{{i}{j}} \bm{\phi}_{a,j}
  - \sum_i
  \frac{2c_2}{c_1} 
  \left(
  \left\lvert 
  \bm{\phi}_{A,i}
  \right\rvert^2
  +
  \left\lvert 
  \bm{\phi}_{B,i}
  \right\rvert^2
  \right)
  \notag\\
  &\hspace{30pt}
  + \sum_i \left[
    u \left(
    \left\lvert 
    \bm{\phi}_{A,i}
    \right\rvert^2
    +
    \left\lvert 
    \bm{\phi}_{B,i}
    \right\rvert^2
    \right)^2
    + \lambda
    \left(
    \bm{\phi}_{A,i}\cdot\bm{\phi}_{B,i}
    \right)^2\right].
  \label{eq:app:4}
\end{align}
Here, one can verify that 
the coefficient 
$
u = 2\left({c_2}/{c_1}\right)^2 - {2c_4}/{c_1} 
$
is positive and 
$
\lambda = {-8c_4}/{c_1} 
$
is negative.
It is straightforward to rewrite the quadratic terms in the form given in Eq.~\eqref{eq:GLW}.

\section{%
  \label{app:review of RG}
  REVIEW OF THE RG TREATMENT ON THE EFFECTIVE HAMILTONIAN
}
In this appendix, we derive the one-loop RG flow equations of $H_{\text{eff}}$
[Eq.~\ref{eq:GLW}]. The final result was first presented by Aharony~\cite{Aharony1975critical} in the
rather different context of amorphous magnets. For completeness, we will consider the generalization to O($N$) of our
O(3) invariant Hamiltonian  $H_{\text{eff}}$ ($N$ is the number of components of each spin).

We start by analyzing the stability of the so called decoupled fixed point (DFP).
This is the Wilson-Fisher fixed point with O($N$) symmetry and 
$u^{\ast} = \epsilon/\left[8\left(N+8\right)\right] + O\left(\epsilon^2\right)$, $u_{AB}^{\ast} = \lambda^{\ast} = 0$.
The reason we are interested in this fixed point is two-fold.
In the first place, the values of the bare coupling $\lambda$ or $u_{AB}$ can be very small for some frustrated magnets such as the $J_1$-$J_2$ model with $J_2 \gg J_1$ or the quasi-2D bct lattice model.~\cite{Sebastian2006dimensional,Batista2007geometric,Schmalian2008emergent,Kamiya2009finite}
In the second place, we can discuss the stability of the DFP very accurately using a non-perturbative scaling argument~\cite{Aharony1975critical,Kamiya2009finite,CalabreseArXivcritical} because the 
sublattices are trivially decoupled and the O($N$) fixed point has been studied very extensively (there are very
accurate estimations of the corresponding exponents).

The stability of our DFP is determined by the scaling dimensions of $u_{AB}$ and $\lambda$, which can be obtained from the two-point correlators of the conjugate scaling operators: 
First, 
\begin{align}
  &\left\langle
  \lvert\bm{\phi}_{A}(x_1)\rvert^2 \lvert\bm{\phi}_{B}(x_1)\rvert^2\,\,
  \lvert\bm{\phi}_{A}(x_2)\rvert^2 \lvert\bm{\phi}_{B}(x_2)\rvert^2
  \right\rangle_{D}
  \notag
  \\
  &\hspace{10pt}
  =\left\langle
  \lvert\bm{\phi}_{A}(x_1)\rvert^2 \lvert\bm{\phi}_{A}(x_2)\rvert^2
  \right\rangle_{D}
  \left\langle
  \lvert\bm{\phi}_{B}(x_1)\rvert^2 \lvert\bm{\phi}_{B}(x_2)\rvert^2
  \right\rangle_{D}
  \notag
  \\
  &\hspace{10pt}
  \propto\lvert x_1 - x_2\rvert^{-4x_{t}}.
  \label{eq:correlator1}
\end{align}
Here the average $\left\langle\cdot\right\rangle_{D}$ is taken under the condition $u_{AB} = \lambda = 0$, and
$x_t = d - 1/\nu$ is the scaling dimension of the energy-density operator at the 3D-O($N$) DFP.
Equation~\eqref{eq:correlator1} shows that the scaling dimension of 
$\lvert{\bm{\phi}}_{A}\rvert^2 \lvert{\bm{\phi}}_{B}\rvert^2$
is simply $2x_{t}$ and thus the RG eigenvalue is
\begin{align}
  y_{u_{AB}}^{(D)} = d - 2x_{t} = \frac{2-d\nu}{\nu} = \frac{\alpha}{\nu},
  \label{eq:yu}
\end{align}
where we have used the hyperscaling relation $\alpha = 2-d\nu$.
Because the specific-heat exponent $\alpha$ of the 3D-O($N$) models is known to be negative for $N \geq 2$,
we conclude that $y_{u_{AB}}^{(D)} < 0$, i.e., the DFP is stable against the $u_{AB}$ term.

Now we discuss the relevance of the $\lambda$ term.
By introducing the traceless symmetric quadrupolar tensors
$
Q^{\mu\nu}_{a} 
= \phi^{\mu}_{a} \phi^{\nu}_{a} 
- N^{-1}\delta^{\mu\nu}
\lvert{\bm{\phi}_{a}}\rvert^2
$ 
($a = A, B$),
we can decompose the $\lambda$ term in the following way:
\begin{align}
  \left({\bm{\phi}}_{A}\cdot{\bm{\phi}}_{B}\right)^2
  = Q_{A}^{\mu\nu} Q_{B}^{\nu\mu} 
  + \frac{1}{N} \lvert{\bm{\phi}}_{A}\rvert^2 \lvert{\bm{\phi}}_{B}\rvert^2.
\end{align}
By using the O($N$) invariance of the decoupled Hamiltonian we obtain:
\begin{align}
  &
  \left\langle
  \left({\bm{\phi}}_{A}\cdot{\bm{\phi}}_{B}\right)^2\left(x_1\right)\,\,
  \left({\bm{\phi}}_{A}\cdot{\bm{\phi}}_{B}\right)^2\left(x_2\right)
  \right\rangle_{D}
  \notag
  \\
  & \hspace{10pt} =
  \left\langle 
  Q_{A}^{\mu\nu}(x_1) Q_{A}^{\kappa\lambda}(x_2)
  \right\rangle_{D}
  \left\langle
  Q_{B}^{\nu\mu}(x_1) Q_{B}^{\lambda\kappa}(x_2)
  \right\rangle_{D}
  \notag
  \\
  & \hspace{10pt}+ N^{-2}
  \left\langle
  \lvert\bm{\phi}_{A}(x_1)\rvert^2 \lvert\bm{\phi}_{A}(x_2)\rvert^2
  \right\rangle_{D}
  \left\langle
  \lvert\bm{\phi}_{B}(x_1)\rvert^2 \lvert\bm{\phi}_{B}(x_2)\rvert^2
  \right\rangle_{D}
  \notag
  \\
  & \hspace{10pt} 
  = \frac{C_{{Q}{Q}}}{\lvert x_1 - x_2\rvert^{4x_{Q}}} + \frac{C_{{t}{t}}}{\lvert x_1 - x_2\rvert^{4x_{t}}}.
  \label{eq:two-point-biquadratic-coupling}
\end{align}
Here $C_{{Q}{Q}}$ and $C_{{t}{t}}$ are nonzero coefficients and $x_{Q}$ is the scaling dimension of the quadrupolar order parameter.
Since the second term is irrelevant at the DFP, 
the scaling dimension of the $\lambda$ term is equal to $2x_{Q}$.
By defining $y_{Q} = d - x_{Q}$, we obtain
\begin{equation}
  y_{\lambda}^{(D)} = d - 2x_{Q} = 2y_{Q} - d.
\end{equation}
Reference~\onlinecite{Calabrese2005harmonic} provides estimates of $y_{Q}$ (denoted as $y_2$ there) for $N = 2, 3, 4, 5$ and $16$.
The value of $y_{Q} = 2$ for $N \to \infty$ is also provided.
In all of these cases we find $y_{\lambda}^{(D)} > 0$, meaning that \textit{the DFP is unstable in the presence of the $\lambda$ term.}

It is then natural to ask whether a stable fixed point exists in the proximity of the unstable DFP.
In the following, we show the results obtained by expanding around the Gaussian fixed
point in $4 - \epsilon$ dimensions to $O(\epsilon)$.
Here, we use the notation introduced by Cardy~\cite{Cardy1996scaling} and 
derive the flow equations to $O(\epsilon)$ by 
applying the operator-product expansion (OPE) method.
We assume that the operators are normalized in such a way that
$
\left\langle
\phi_{a}^{\mu}\left(x_1\right)
\phi_{b}^{\nu}\left(x_2\right)
\right\rangle
= \delta_{ab}\delta_{\mu\nu}
\left\lvert
x_1 - x_2
\right\rvert^{-\left(d-2\right)}
$
at the Gaussian fixed point ($a, b = A, B$ and $1\le \mu,\nu\le N$) and that they are normal-ordered in a sense described in Ref.~\onlinecite{Cardy1996scaling}. 
The following OPE's are sufficient to construct
the RG equations:
\begin{align}
  \psi_{r} \cdot \psi_{r} 
  & = 4N + 4 \psi_{r} + \psi_{u} + 2 \psi_{u_{AB}},
  \displaybreak[0]
  \notag\\[+2pt]
  \psi_{r} \cdot \psi_{u} 
  & = 4\left(N+2\right) \psi_{r} + 8 \psi_{u},
  \displaybreak[0]
  \notag\\[+2pt]
  \psi_{r} \cdot \psi_{u_{AB}} 
  & = 4N \psi_{r} + 8 \psi_{u_{AB}},
  \displaybreak[0]
  \notag\\[+2pt]
  \psi_{r} \cdot \psi_{\lambda} 
  & = 2 \psi_{r} + 8 \psi_{\lambda},
  \displaybreak[0]
  \notag\\[+2pt]
  \psi_{u} \cdot \psi_{u} 
  & = 24N^2 + 32\left(N+2\right) \psi_{r} + 8\left(N+8\right) \psi_{u},
  \displaybreak[0]
  \notag\\[+2pt]
  \psi_{u} \cdot \psi_{u_{AB}} 
  & = 8\left(N+2\right) \psi_{u_{AB}},
  \displaybreak[0]
  \notag\\[+2pt]
  \psi_{u} \cdot \psi_{\lambda} 
  & = 8 \psi_{u_{AB}} + 16 \psi_{\lambda},
  \displaybreak[0]
  \notag\\[+2pt]
  \psi_{u_{AB}} \cdot \psi_{u_{AB}} 
  & = 4N^2 + 8N \psi_{r} + 2N \psi_{u} + 16 \psi_{u_{AB}},
  \displaybreak[0]
  \notag\\[+2pt]
  \psi_{u_{AB}} \cdot \psi_{\lambda} 
  & = 4N +  8 \psi_{r} + 2 \psi_{u} + 16 \psi_{\lambda},
  \displaybreak[0]
  \notag\\[+2pt]
  \psi_{\lambda} \cdot \psi_{\lambda}
  & = 4N^2 + 4\left(N+1\right) \psi_{r} + 2 \psi_{u} + 4 \psi_{u_{AB}} 
  \displaybreak[0]
  \notag\\
  &\hspace{80pt}
  + 4\left(N+2\right) \psi_{\lambda}. 
  \label{eq:OPE}
\end{align}
Here, 
$
\psi_{r} \equiv \lvert{\bm{\phi}}_{A}\rvert^2 + \lvert{\bm{\phi}}_{B}\rvert^2
$,
$
\psi_{u} \equiv \lvert{\bm{\phi}}_{A}\rvert^4 + \lvert{\bm{\phi}}_{B}\rvert^4
$,
$
\psi_{u_{AB}} \equiv \lvert{\bm{\phi}}_{A}\rvert^2\,\lvert{\bm{\phi}}_{B}\rvert^2
$ 
and
$
\psi_{\lambda} \equiv \left(\bm{\phi}_{A} \cdot \bm{\phi}_{B}\right)^2
$
are short-hand notations for the scaling operators.
The RG flow equations to $O(\epsilon)$ are entirely determined by these
OPE coefficients:~\cite{Cardy1996scaling}
\begin{align}
  \frac{{d}{r}}{{d}{l}} &= 2 r - 8(N + 2) r u - 4 r \lambda- 4 N r u_{AB} -\dots, 
  \label{eq:one-loop-r}
  \\
  \frac{{d}{u}}{{d}{l}} &= \epsilon u - 8(N + 8) u^2 - 2 \lambda^2 - 4 u_{AB} \lambda 
  \notag\\
  &\hspace{90pt}
  - 2N u_{AB}^2 -\dots,
  \label{eq:one-loop-u}
  \\
  \frac{{d}{u_{AB}}}{{d}{l}} &= \epsilon u_{AB} - 16 u \lambda - 16(N+2) u u_{AB} - 4\lambda^2 
  \notag\\
  &\hspace{90pt}
  - 16u_{AB}^2 -\dots, 
  \label{eq:one-loop-uAB}
  \\
  \frac{{d}{\lambda}}{{d}{l}} &= \epsilon \lambda - 32u\lambda - 4(N+2) \lambda^2 - 32u_{AB} \lambda -\dots, 
  \label{eq:one-loop-lambda}
\end{align}
where we assume that the fixed points of physical interest are located in the region where
$r = O(\epsilon^2)$, $u = O(\epsilon)$, $u_{AB} = O(\epsilon)$ and $\lambda = O(\epsilon)$.

Let us first discuss the physically relevant cases $N = 2, 3$.
The fixed points to $O(\epsilon)$ for $N = 2$ are as follows:
\begin{itemize}
\item Gaussian fixed point: $\left(u, u_{AB}, \lambda\right) = \left(0, 0, 0\right)$
\item XY DFP: $\left(u, u_{AB}, \lambda\right) = \left(\epsilon/80, 0, 0\right)$
\item O(4)-like: $\left(u, u_{AB}, \lambda\right) = \left(\epsilon/96, \epsilon/48, 0\right)$
\item $\left(u, u_{AB}, \lambda\right) = \left(\epsilon/160, 3\epsilon/80, -\epsilon/40\right)$
\item $\left(u, u_{AB}, \lambda\right) = \left(\epsilon/160, \epsilon/80, \epsilon/40\right)$
\end{itemize}
The fixed points to $O(\epsilon)$ for $N = 3$ are as follows:
\begin{itemize}
\item Gaussian fixed point: $\left(u, u_{AB}, \lambda\right) = \left(0, 0, 0\right)$
\item O(3) DFP: $\left(u, u_{AB}, \lambda\right) = \left(\epsilon/88, 0, 0\right)$
\item O(6): $\left(u, u_{AB}, \lambda\right) = \left(\epsilon/112, \epsilon/56, 0\right)$
\item $\left(u, u_{AB}, \lambda\right) = \left(3\epsilon/272, \epsilon/136, 0\right)$
\item $\left(u, u_{AB}, \lambda\right) = \left(\epsilon/136, \epsilon/68, \epsilon/68\right)$
\item $\left(u, u_{AB}, \lambda\right) = \left(\epsilon/176, \epsilon/88, \epsilon/44\right)$
\end{itemize}
The most important conclusion is that \textit{none of these fixed points is stable to $O(\epsilon)$}. 
Therefore, this simple RG calculation suggests that the  biquadratic coupling $\lambda$ between the two $O(N)$
subsystems leads to a first-order phase transition.~\cite{Aharony1975critical}
It is interesting to note that, in contrast to the result obtained by directly evaluating the correlation function, the one-loop expansion indicates that $u_{AB}$ is a relevant perturbation at the DFP for $N < 4$ and $d < 4$: 
\begin{equation}
  y_{u_{AB}}^{(D)} = \left(4-N\right)\epsilon/\left(N+8\right) + O(\epsilon^2).
  \label{eq:one-loop-yu}
\end{equation}
Note that $y_{u_{AB}}^{(D)}$ is positive to $O(\epsilon)$ for $N < 4$ and $d < 4$, while our Eq.~\eqref{eq:yu} shows
 that $y_{u_{AB}}^{(D)} < 0$ for $N \geq 2$ in $d = 3$.
This discrepancy must be eliminated by the higher-order terms of the $\epsilon$ expansion.
Although the concomitant change that will appear in the RG flow structure is
unclear, numerical studies of microscopic Hamiltonians, such as the Ising-O(3)
model in the present work or the coupled XY model in Ref.~\onlinecite{Kamiya2010crossover}, confirm the absence of a stable fixed point. 
The model does not have a stable fixed point in the region $\lambda < 0$ even
for larger values of $N$.~\cite{Aharony1975critical}


\bibliography{references}

\end{document}